\documentclass{article}

\usepackage[final]{neurips_2023_ml4ps}

\usepackage[utf8]{inputenc} 
\usepackage[T1]{fontenc}    
\usepackage{hyperref}       
\usepackage{url}            
\usepackage{booktabs}       
\usepackage{amsfonts}       
\usepackage{nicefrac}       
\usepackage{microtype}      
\usepackage{xcolor}         
\usepackage{natbib}
\usepackage[pdftex]{graphicx}
\setcitestyle{square,sort,comma,numbers}
\title{Combining astrophysical datasets with CRUMB}

\author{%
  Fiona A.~M.~ Porter \& Anna M.~M.~ Scaife\thanks{The Alan Turing Institute, 96 Euston Rd, London, UK \texttt{a.scaife@turing.ac.uk}}\\
  Jodrell Bank Centre for Astrophysics \\
  University of Manchester \\
  \texttt{\{fiona.porter, anna.scaife\}@manchester.ac.uk} \\
  }
  
\begin{document}

\maketitle

\begin{abstract}
  At present, the field of astronomical machine learning lacks widely-used benchmarking datasets; most research employs custom-made datasets which are often not publicly released, making comparisons between models difficult. In this paper we present CRUMB, a publicly-available image dataset of Fanaroff-Riley galaxies constructed from four ``parent'' datasets extant in the literature. In addition to providing the largest image dataset of these galaxies, CRUMB uses a two-tier labelling system: a ``basic'' label for classification and a ``complete'' label which provides the original class labels used in the four parent datasets, allowing for disagreements in an image's class between different datasets to be preserved and selective access to sources from any desired combination of the parent datasets. 
\end{abstract}

\section{Introduction}

The field of astronomy is entering the era of Big Data astrophysical surveys, with peta- or exascale volumes of data anticipated with the advent of instruments such as the Square Kilometre Array \citep{scaife2020big}. The traditional method of classifying sources – visual inspection by astronomers – will clearly be impractical for these surveys, and a natural solution is to turn instead to machine learning classification. However, astronomical machine learning research does not at present have standard benchmarking datasets; instead, it is common for researchers to develop their own datasets for their publications, most of which are not made publicly available, making it difficult to reproduce results or compare the performance of different techniques. While this is being addressed in part by the release of large labelled datasets by e.g. \citep{walmsley2022galaxy}, an additional measure that can be taken is to build combined datasets by identifying compatible datasets in the existing literature and integrating them together. This work will discuss the considerations needed when combining astronomical datasets, the methods used to construct a merged dataset and the applications of the resulting dataset. 

\section{Dataset construction}

\setcounter{footnote}{0} 

\begin{figure}
    \centerline{
        \includegraphics[width=0.32\textwidth]{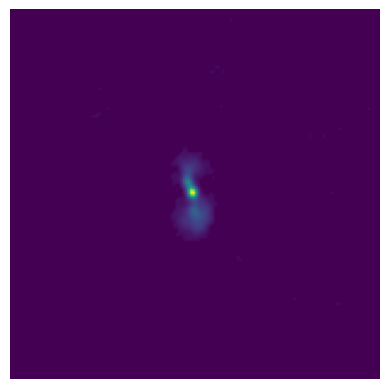}\quad \includegraphics[width=0.32\textwidth]{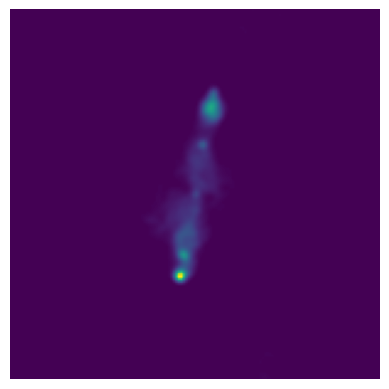}\quad
        \includegraphics[width=0.32\textwidth]{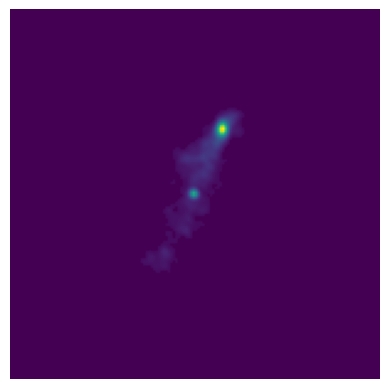}
        }
    \centerline{(a) \hskip 0.32\textwidth (b) \hskip 0.32\textwidth (c)}
    \centering
    \caption{Examples of FRI (a), FRII (b) and hybrid (c) morphology from CRUMB. (a) is found in FR-DEEP and AT17; (b) is found in MB and FRDEEP; (c) is found in MB-Hyb only.}
    \label{fr morphologies}
\end{figure}

For this work we chose to focus upon datasets of Fanaroff-Riley (FR) galaxies \citep{fanaroff1974morphology}, which are active galactic nuclei (AGN) which emit very brightly at radio wavelengths. These galaxies are classified as either FRI, FRII or hybrid based on their morphology (see Figure~\ref{fr morphologies}), with said morphologies being informative of the properties of the host AGN engines and surrounding environments. Despite a number of publications using this population for machine learning \citep[e.g.][]{aniyan2017classifying, ma2019machine, scaife2021fanaroff, mohan2022quantifying}, existing publicly-available datasets of FR galaxies are comparatively small; the largest at the time this work began, the MiraBest dataset, contains only 1256 sources \citep{porter2023mirabest}. Creating a combined dataset using MiraBest and any other suitable datasets of FR galaxies was hence seen to have two benefits: it would serve as an excellent test case for dataset merging and also provide an appreciably larger population of FR galaxies within a single dataset for the astronomical community. We chose to dub this combined dataset Collected Radiogalaxies Using MiraBest (CRUMB)\footnote{\url{https://doi.org/10.5281/zenodo.7948346}}. 

\subsection{Selecting datasets to merge}

When attempting to merge astronomical machine learning datasets, a key consideration is whether they draw their image data from surveys with sufficiently similar properties. Astrophysical sources can have markedly different morphological characteristics at different wavelengths, so for consistency in feature detection all image data should be drawn from surveys of the same wavelength. As well as this, the appearance of sources naturally varies when observed by instruments with different resolution; a survey with higher angular resolution can reveal extended structure in what would appear to be point sources when observed by a survey with lower angular resolution, but may ``resolve out'' (and fail to detect) faint extended emission that is visible in the lower-resolution survey. Again, this can result in inconsistencies in the appearance of sources of the same class, as they are effectively being shown at different spatial scales.

A natural solution to these concerns is to simply make use of datasets which were constructed using the same survey. In radio astronomy, the VLA FIRST survey \citep{becker1994vla} is a relatively common choice for dataset construction, being large and readily accessible; it provides coverage of a large area of the northern sky at the commonly-used frequency of 1.4 GHz, at a comparably high resolution of 1.8 arcseconds per pixel, and its data are publicly available. Limiting our search to extant image datasets of FR galaxies using FIRST data, two datasets were identified as being suitable for combination with the MiraBest dataset (hereafter MB) \citep{porter2023mirabest} and its supplemental hybrid dataset (hereafter MB-Hyb): the FR-DEEP dataset \citep{tang2019transfer} and the unnamed dataset of \citep{aniyan2017classifying} (hereafter AT17).

\subsection{Cross-matching sources}

Astronomical datasets are often derive their source lists from catalogues of the desired source classes, which provide class labels and coordinates for each source's centre; however, different catalogues may disagree on the exact location of the centre, leading to multiple sets of coordinates being used for the same source. Since the four ``parent'' datasets used to create CRUMB are derived from a total of six different catalogues, it was necessary to cross-match their sources to remove any duplicates.

To cross-match, the coordinates of all sources in the four datasets were compared to one another, with pairs of sources being flagged if they were less than 270 arcseconds (the width of one image) apart. Flagged sources were visually inspected to determine whether the two sets of coordinates corresponded to the same source; if so, we retained the best match to the visual centre as the ``preferred'' coordinates and discarded any others. Using this method, a total of 2100 unique sources were identified from the 2731 sets of coordinates in the parent datasets. Seven of these sources were additionally found to be poorly centred and were realigned accordingly.

\subsection{Data processing}

Image data for each source were collected from the FIRST survey \citep{becker1994vla} using SkyView Virtual Telescope \citep{mcglynn1994skyview}, which was used to produce a $300\times300$ pixel (540") image centred upon the source in FITS format. Radio noise was removed by using the \verb|astropy| package's \verb|sigma_clipped_stats| function \citep{astropy2022astropy} to identify all pixels with value $< 3\sigma$ and setting them to zero, at which point the images were cropped to their final size of $150\times150$ pixels. Next, a circular mask of diameter 150 pixels was applied to remove any bright background sources present around the edge of the images. Finally, the source data were normalised by determining the minimum and maximum pixel flux values for each image and scaling each image pixel as follows:

\begin{equation}\label{eq: gal entropy}
    Normalised\,pixel\,value = 255 \times \frac{Pixel\, value - minimum\, flux}{Maximum\, flux - minimum\, flux}.
\end{equation}

Here, the factor of 255 allows for maximum dynamic range for PNG format, which all images were saved as. While keeping the images in FITS format would have allowed for a higher dynamic range to be preserved, this format is not common outside of astronomy and the resulting file sizes were found to be two orders of magnitude larger than the same source in PNG format; PNG was hence preferred both for accessibility to the broader AI community and reduced file size. 

\subsection{Creating class labels}
\label{labels}

Each of the datasets used in the creation of CRUMB makes use of a different set of class labels for its sources (see Table~\ref{class-labels}); MB is the most complex, having a total of ten classes to mark both morphological subclass and certainty in classification, while AT17 has three classes and both FR-DEEP and MB-Hyb have two. Determining a unified labelling system is hence somewhat complex, particularly as the labelling systems of MB and AT17 are somewhat incompatible, with bent-tailed sources being considered a subclass of FRIs by MB but an entirely separate class by AT17. Additionally, of the 541 sources which were present in more than one dataset, 64 were found to have disagreements in label between the different datasets and 15 had labels which were entirely contradictory; clearly, there is disagreement event amongst experts about how some sources should be labelled, and selecting a single ``correct'' label is not trivial. 

\begin{table}
  \caption{Class labels used in each of the parent datasets. ``WAT'' and ``HT'' are bent-tail morphologies treated as FRIs by MB, equivalent to ``bent'' sources in AT17; ``DD'' are double-double FRIIs.}
  \label{class-labels}
  \centering
  \begin{tabular}{cllll}
    \toprule
    \textbf{Label} & \multicolumn{1}{c}{\textbf{MB}} & \multicolumn{1}{c}{\textbf{FR-DEEP}} & \multicolumn{1}{c}{\textbf{AT-17}} & \multicolumn{1}{c}{\textbf{MB-Hyb}} \\ 
    \midrule
    0 & Confident FRI & FRI & FRI & Confident hybrid \\
    1 & Confident WAT & FRII & FRII & Uncertain hybrid \\
    2 & Confident HT &  & Bent &  \\
    3 & Uncertain FRI &  &  &  \\
    4 & Uncertain WAT &  &  &  \\
    5 & Confident FRI &  &  &  \\
    6 & Confident DD &  &  &  \\
    7 & Uncertain FRII &  &  &  \\
    8 & Confident hybrid &  &  &  \\
    9 & Uncertain hybrid &  &  &  \\
    \bottomrule
  \end{tabular}
\end{table}

Rather than attempting to merge the disparate labelling systems, we make use of the alternative approach of a two-tier labelling system: each image is given both a ``basic'' label of either FRI, FRII or hybrid and a ``complete'' label which contains its class labels in all of the parent datasets. This compromise allows for CRUMB to offer both an easy-to-understand labelling system for users who simply wish to train models using astronomical data and a secondary, more complex system for users who wish to reproduce results using a specific parent dataset or limit the sources they use based on requirements such as e.g. removing any sources with label ambiguity. 

CRUMB uses three classes for its basic labels - FRI (0), FRII (1) or hybrid (2) - and contains a total of 1006 FRIs, 997 FRIIs and 97 hybrids. All sources labelled as ``bent'' by AT17 were folded into the FRI class to align with the primary class labels of MB; if users wish to instead treat bent-tailed sources as a distinct class, this can be done by applying logical filtering to the complete labels (see Section ~\ref{label access}). Each complete label is a vector with four entries and provides the label for the source in each of the parent datasets per their respective labelling schemes, in the order MB, FR-DEEP, AT17, MB-Hyb. If a source is not present in a particular dataset, it is denoted ``-1''. As an example, a source with MB label ``0'' (FRI) and AT17 label ``2'' (bent source) would have a complete label of ``[0, -1, 2, -1]''. In this way, multiple contradictory labels can be preserved.

\subsection{Building the batched dataset}

CRUMB was designed as a \textit{batched dataset} \citep{masters2018revisiting}, capable of loading each data batch into memory sequentially rather than loading the entire dataset at once, making it practical to use on machines with limited memory such as personal laptops. It was decided to split the dataset into seven batches of 300 images each, with one batch being reserved as a test set. 

There is a significant class imbalance in CRUMB, with both the FRI and FRII classes containing around ten times as many sources as the hybrid class. Randomly assigning images to each batch might hence result in significant variation in the number of hybrid sources per batch, and it was deemed that an underpopulation of hybrids in the test set in particular might negatively affect model evaluation. To ensure this did not occur, hybrid sources were evenly distributed between all batches before FRIs and FRIIs were added, resulting in a minimum of 13 hybrids in each batch. With this done, the image data and both sets of labels were collected for each batch and used to build the final dataset, which was then made publicly accessible on Zenodo \citep{porter2023crumb}. 

\section{Using CRUMB}

In addition to being (to our knowledge) the largest publicly-available machine learning dataset of FR galaxies, CRUMB was designed with the intent of being adaptable to its users' needs and capable of being built on further should more suitable data be found. This section will discuss how to access various features of CRUMB and the possibility of its expansion. 

\subsection{Accessing labels and subclasses}
\label{label access}

CRUMB is designed for use with Python machine learning packages such as Pytorch \citep{paszke2019pytorch} and Keras \citep{chollet2015keras}, and the dataset is provided with a Python class using a structure inherited from Pytorch's \texttt{data.Dataset} class. 

As well as the standard \textit{targets} method providing the class for each source, CRUMB offers the \textit{complete\_labels} method to access the complete class label for each of the parent surveys (see Section~\ref{labels}). Using \textit{complete\_labels}, it is possible to logically filter CRUMB to include only sources which match user criteria for either training or testing. This may be done by loading the full dataset, identifying desired source properties (e.g. labelled as "bent" by AT17), applying logical operations to \textit{complete\_labels} to identify these sources (e.g. complete\_labels[2] == 2), and passing the resulting indices to a subloader to use only those images; this method is demonstrated on CRUMB's github page\footnote{\url{https://github.com/fmporter/CRUMB}}.

Additionally, a number of subclasses have been created for subsets that we expect users might find useful. These include loaders for each of the parent datasets, combined loaders to access MB and MBHyb simultaneously or include no sources in MB or MBHyb, and a four-class loader which treats bent sources as a separate class from FRIs. All parent dataset loaders include a flag to default to either the basic label or the dataset's original labelling scheme, and all combination loaders include a flag to specify which dataset's labels should be used in the case of a contradiction. Additional subclasses may be easily constructed using these loaders as templates, making CRUMB flexible and customisable to its users' needs.

\subsection{Building on CRUMB}

All sources in CRUMB are fully traceable to both their parent datasets via the \textit{complete\_labels} method and the J2000 coordinates of the pictured source using the \textit{filenames} method. This means that it would be possible to process additional image data in the same manner as CRUMB (as detailed both here and in \cite{porter2023mirabest}), cross-match sources and append an additional column to the complete label accordingly to expand upon CRUMB if desired. Similarly, because of the sources' traceability it would be possible to produce a dataset of the same sources using survey data at a different frequency or resolution if desired for e.g. transfer learning applications \citep{tang2019transfer}. We intend to add more sources to CRUMB if other compatible datasets are created in the future, maintaining it as a single dataset from which all previous versions can be obtained by filtering the complete labels. 

\section{Conclusions}

Astronomical machine learning datasets are, at present, often not published along with the papers they were built for, making it more difficult to draw comparisons between techniques from different publications. To improve reproducibility of our research and offer a resource in the form of a dataset for both astronomers who want to investigate machine learning and for computer scientists considering using astronomical data, we have presented CRUMB, a dataset of FR galaxies designed to combine four existing astronomical machine learning datasets in a flexible, accessible and reproducible manner. CRUMB was purposefully designed to be straightforward to expand as new data becomes available, so the authors anticipate its expansion in the future to ensure it remains a useful resources when comparing studies of FR galaxies.

\begin{ack}

FMP and AMS gratefully acknowledge support from Alan Turing Institute AI Fellowship EP/V030302/1. FMP would also like to thank Hongming Tang and Kshitij Thorat for sharing the source lists for their datasets.

\end{ack}



\bibliography{neurips_2023}






\end{document}